# DIY the Integrated Climate Model and its computational performance

WANG Pengfei[1,2]

[1]Center for Monsoon System Research, Institute of Atmospheric Physics, Chinese Academy of Sciences, Beijing 100190, China

[2]State Key Laboratory of Numerical Modeling for Atmospheric Sciences and Geophysical Fluid Dynamics, Institute of Atmospheric Physics, Chinese Academy of Sciences, Beijing 100029, China

Email: wpf@mail.iap.ac.cn

**Abstract** This article describes the software engineering framework and computation performance of a global climate system model which helps the user to understand the step-by-step technical to DIY(do it yourself) a climate model by your own. The model integrates ECHAM5 and NEMO2.3 using OASIS3 as the coupler. The program skill of the Integrated global Climate Model (ICM) is demonstrated here, including the porting of NEMO into the COSMOS framework, the organization of variable exchange, and component model communication synchronization. We describe how we successfully fixed some bugs in the component models and detail the new code and scripts that were added to ICM. In particular, an improvement of ICM's coding is that we enabled it to perform perfect restart runs, which is an important feature that was not implemented in the original version of NEMO2.3. ICM is designed as a model hybrid for both research and operational applications. Its scripts manage and utilize numerical simulations including data preparation, namelist file control, CPU allocation, job submission, job restarting, post-data combination and plotting. ICM is now successfully set up and runs at Dawning in the CMSR and the Tianhe-1A supercomputer system in the National Supercomputer Center in Tianjin, and the speed benchmark has been obtained. Results indicate that it can simulate the climate at more than 100 model years per day (maximum of 175 yrs/d), which represent a high level of performance for a fully coupled climate model. Finally, a summary of the 1500-year long-term control experiment using ICM is provided.

## 1. Introduction

Climate modeling is an essential tool that is widely used to provide an understanding of the evolution of climate systems and to make climate predictions. An AOGCM was developed from 2008 by the Center for Monsoon System Research, Institute of Atmospheric Physics (CMSR/IAP). This model integrates version 5 of the Hamburg atmospheric general circulation model [1] and version 2.3 of the Nucleus for European Modeling of the Ocean [2] using version 3 of the Ocean Atmosphere Sea Ice Soil as the coupler [3]. The name of the model is the Integrated Couple Model (ICM). To date, ICM has steadily integrated 1500 years without climate drift in the Dawning and Tianhe-1A (TH-1A) [4] supercomputer system. The model can reproduce a realistic distribution of seasonal SST, with bias of around 1–2°C relative to observations in the tropics. The seasonal cycle of equatorial Pacific SST is one of the common challenges of AOGCMs and is also reasonably simulated, although the westward phase propagation of tropical SST is not accurately described. For details regarding the model's provision of results with respect to the seasonal prediction of EA-WNP climate, readers are referred to Huang et al. [5].

A proposed application of ICM is in the simulation of ENSO dynamics. The dominant interannual pattern of the tropical Pacific, i.e., El Niño, is reproduced with a realistic spatial pattern, variance and period, whereas the inter-decadal variation of tropical Pacific SST is underestimated in the model. The main biases of ICM are the excessive cold tongue associated with deficient precipitation over the central Pacific, as is the case in many other models. ICM reproduces many kinds of atmospheric and oceanic fields and it also supports ENSO initial data and the ensemble generation method. With the help of ICM, we can complete both hindcasts and forecasts of ENSO[6].

The goal of this article is express the software engineering framework and computational performance of ICM to help understanding the step-by-step technical to DIY(do it yourself) a climate model, and we also demonstrate some of its operational skills. The organization of this article is as follows. In section 2 and 3, we generalize the structure of the model code and the coupling fields. In section 4, we describe the way to find out and fix the bugs of components models. We then provide the scripts system to do effective and long-term simulations in section 5. Finally, in section 6 we obtain the computational performance of ICM in TH-1A platform.

**2. The structure of model code**

The first version of ICM is a coupled atmosphere–ocean–sea ice model without flux adjustment. A schematic diagram of ICM is shown in Fig. 1. ICM's atmospheric model is ECHAM5, its oceanic and sea ice model is NEMO 2.3 (hereafter referred to simply as NEMO), and its coupler is OASIS3. ECHAM and NEMO were chosen as the atmospheric and oceanic components because they possess high skill in simulating global general circulation and EA-WNP regional climate. Indeed, other existing coupled models that employ them show strong stability and usability. OASIS3 is a universal coupler and has been used in many coupled models[3]. The fields exchanged between atmospheric and oceanic model in ICM is similar to the Kiel Climate Model (KCM)[7], CMCC-INGV[8] and SINTEX [9, 10].

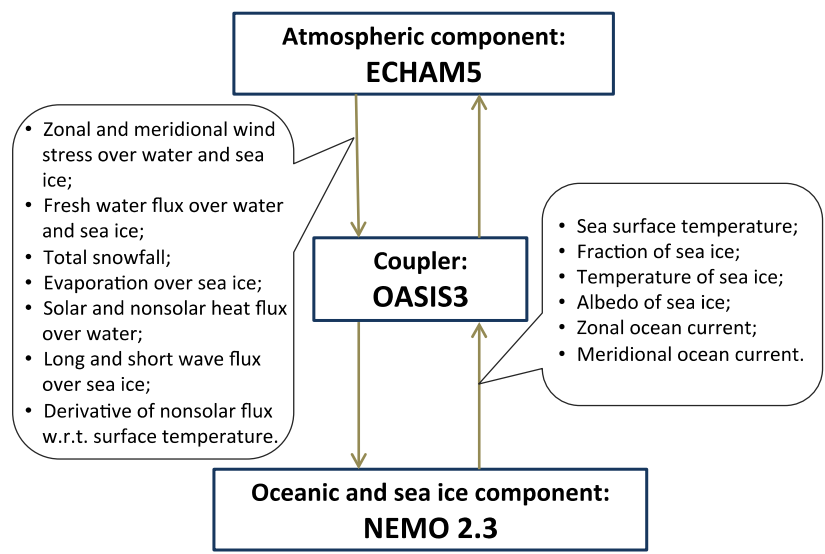

Figure 1.    Schematic diagram of ICM.

The program coding framework of ICM follows that of the COSMOS [11] Earth System Model (now renamed as MPI-ESM), which was originally designed to investigate the global marine carbon cycle over long timescales [12]. We set up and run the COSMOS model in a Linux cluster. The compiler option and library was different to that for a vector computer, and so a patch named cosmos-1.0.0-csc.patch (provided by the CSC – IT Center for Science Ltd, which is administered by the Ministry of Education, Science and Culture in Finland) was necessary. After patching this file, the COSMOS system could be compiled successfully with some small modifications, such as the NETCDF and MPI paths.

The COSMOS-1.0 system comprises the ECHAM5 and MPI-OM models. Before embarking on the next stage of development for ICM, we should run the original COSMOS model to test the compilation environments and MPI communication of the cluster to ensure they are all working satisfactorily. The correct computer environment and platform is beneficial to reducing the difficulties involved in further model porting.

In terms of a model's software engineering, a well written technical document or demo codes are sometimes more valuable. Useful documents for porting a new model component and integrating the field exchanges between new components and original ones in COSMOS are provided by Valcke [3, 13]. Here, we emphasize two further technical reports other than journal publications: the first is an introduction to SINTEX-F by A. Caubel (from IPSL) and E. Maisonnave (from CERFACS, Jun 2004); and the second is actually a series of reports by the IPSL-ES, but especially a French report written by Eric Maisonnave named "Implémentation du modèle couplé ECHAM-NEMO LIM (30 et 300 niveaux) sur le Earth Simulator du 9 au 28 Août 2006". By referring to these documents, we can keep abreast of the successful work of SINTEX-F and save much time when considering specific work other than the main coupling strategies.

Figure 2a demonstrates the main source directory structure of ICM The atmospheric and oceanic model components are located in the 'src/mod' directory.

```
|-- src                    (a)      nemo                (b)
|   |-- include                     |-- C1D_SRC
|   |   |-- make_dummies            |-- LIM_SRC
|   |   |-- make_dummies_libs       |-- NST_SRC
|   |-- lib                         |-- OPA_SRC
|   |   |-- anaisg                  |   |-- DIA
|   |   |-- anaism                  |   |-- DOM
|   |   |-- blas                    |   |-- DTA
|   |   |-- clim                    |   |-- DYN
|   |   |-- fscint                  |   |-- FLO
|   |   |-- ioipsl                  |   |-- IOM
|   |   |-- lapack                  |   |-- LDF
|   |   |-- mpp_io                  |   |-- OBC
|   |   |-- psmile                  |   |-- SBC
|   |   |-- scrip                   |   |-- SOL
|   |   `-- support                 |   |-- TRA
|   `-- mod                         |   |-- TRD
|       |-- echam5                  |   `-- ZDF
|       |-- nemo                    |-- include
|       `-- oasis3                  `-- src
```

Figure 2.    Source tree of ICM: (a) global source; (b) NEMO source.

**2.1 Porting ECHAM5 into ICM**

ECHAM5, the fifth version of the European Centre for Medium-Range Weather Forecasts (ECMWF) model developed by the Max Planck Institute for Meteorology, is a spectral model with state-of-the-art physics [1, 14]. ECHAM5 has shown good performance in reproducing the climatological precipitation and interannual pattern of the East Asian summer monsoon [15]. The low-resolution version (T31; horizontal resolution: 3.75° × 3.75°; vertical resolution: 19 levels) of ECHAM5 is used in the first version of ICM. ECHAM5 has developed some higher-resolution versions, and thus the model resolution can be easily improved in ICM. For more details regarding ECHAM5, readers are referred to Roeckner [1].

In spite of ICM having been developed to follow the COSMOS framework, and ECHAM5 already being a component model coupled to the MPI-OM ocean model, we still needed to modify much of the Fortran code in ECHAM5 to handle the variable data received from NEMO. More details are provided in section 3.1.

## 2.2 Porting NEMO into ICM

Version 2.3 of NEMO includes version 9 of the Ocean Parallelise (OPA9) ocean general circulation model (OGCM), developed by Institut Pierre-Simon Laplace (IPSL) [2], and (for sea ice) version 2 of the Louvain-la-Neuve Ice Model (LIM2), developed by Louvaine-la-Neuve [16]. The oceanic component, OPA9, is a finite difference OGCM, and the Arakawa C-grid is used to solve the primitive equations. OPA9 uses a global orthogonal curvilinear ocean mesh to move the mesh poles to the Asian and North American continents for avoiding singularity points in the computational domain. The horizontal grid of NEMO uses an ORCA2 grid with horizontal grid numbers of 182 (longitude) × 149 (latitude) (around 2° at high latitudes and an enhanced meridional resolution of 0.5° close to the equator), and 31 vertical levels to a depth of 5250.23 m with 10 levels in the top 100 m. More details on NEMO can be found in Madec [2].

Like the situation with ECHAM5, NEMO was originally coupled to a French atmospheric model, LMDZ-4. Thus, when it was coupled with ECHAM5, we needed to provide patches to handle the data exchange process. An IPSL note[17] describes the NEMO model in IPSL-modipsl style to define the keys and compiler option.

We placed the NEMO source in src/mod/nemo, and the detailed source structure is depicted in Fig. 2b. In the nemo/src/ directory, we linked all necessary sources that are distributed in other directories. Moreover, we generated a new Makefile for these sources by COSMOS's Makefile generating tool[18]. The compilation script is located in the nemo directory, which was ported from the script used for MPI-OM and modified for the model name. Some model-dependent definitions were also added.

The compile scripts for NEMO have two different series: one for generating an uncoupled ocean model, and another for compiling as a coupled model component. As mentioned, the uncoupled NEMO model can be compiled by modipsl style, and we can compare the uncoupled NEMO compilation result through these two different approaches. The input data for the uncoupled NEMO model in ICM is the same as that compiled by modipsl. The data can be download from the IPSL website. The resolution for NEMO is set to ORCA2, with a corresponding grid of 182 × 149 with 31 vertical levels. After confirming that NEMO can simulate normally, we were able to conclude that NEMO had been successfully ported into ICM, at least as an independent model.

The script COMP_nemo_nemo.ksh is the uncoupled NEMO compile script, and we can add -Dkey_xxxxx to enable or disable the NEMO physical process. Table 1 lists the commonly used keys for NEMO, and the keys for the coupled model are listed at the bottom of the table.

**Table 1. Some of the keys used for NEMO.**

| Key | Description | Notes |
| --- | --- | --- |
| -Dkey_vectopt_loop | Vector computer | Not used at Linux |
| -Dkey_vectopt_memory | Vector computer | Not used at Linux |
| -Dkey_orca_r2 | Define the resolution | Options: key_orca_r4 key_orca_r1 key_orca_r05 key_orca_r025 |
| -Dkey_ice_lim | Activate the LIM sea ice model | |
| -Dkey_lim_fdd | Freshwater flux | |
| -Dkey_dynspg_flt | Filtered free surface | |
| -Dkey_diaeiv | Diagnose eddy-induced velocities | |
| -Dkey_ldfslp | Compute the slopes of neutral surface | |
| -Dkey_traldf_c2d | Latitude- and longitude-dependent coefficient | |
| -Dkey_traldf_eiv | Eddy-induced velocity | |
| -Dkey_dynldf_c3d | Latitude-, longitude-, and depth-dependent coefficient | |
| -Dkey_dtatem | Activate climatic SST | Use when uncoupled |
| -Dkey_dtasal | Activate climatic salinity | Use when uncoupled |
| -Dkey_tradmp | Tracer damping | |
| -Dkey_trabbc | Bottom boundary condition | |
| -Dkey_trabbl | Bottom boundary layer | |
| -Dkey_trabbl_dif | Diffusive bottom boundary layer | |
| -Dkey_diahth | Output of 20 isotherms | |
| -Dkey_zdftke | tke 1.5 turbulent closure scheme | |
| -Dkey_zdfddm | Double diffusion mixing | |
| -Dkey_mpp_mpi | Enable MPI in NEMO | |
| **Keys used in the coupled mode** | | |
| -Dkey_coupled | Run NEMO in coupled mode | |
| -Dkey_oasis3 | Use Oasis3 as the coupler | |
| -Dkey_cpl_albedo | Exchange the albedo | |
| -Dkey_cpl_ocevel | Exchange the 'ocevel' variable | |
| -Dkey_root_exchg | Use root process to communicate | |

The actual keys used to compile a coupled NEMO model are:

CPPFLAG=" -Dkey_mpp_mpi -D__Linux -Dkey_trabbl_dif -Dkey_orca_r2 -Dkey_ice_lim -Dkey_lim_fdd -Dkey_dynspg_flt -Dkey_diaeiv -Dkey_ldfslp -Dkey_traldf_c2d -Dkey_traldf_eiv -Dkey_dynldf_c3d -Dkey_dtatem -Dkey_dtasal -Dkey_tradmp -Dkey_trabbc -Dkey_zdftke

-Dkey_zdfddm -Dkey_coupled -Dkey_oasis3 -Dkey_cpl_albedo -Dkey_cpl_ocevel –Dkey_diahth -Dkey_cpl_rootexchg".

These keys can be changed manually, and the clean and rebuild operations are needed to activate the new keys.

**2.3 OASIS in ICM**

The OASIS3 [3] coupler is used to communicate between ECHAM5 and NEMO in ICM. Seventeen variables – including wind stress, surface heat flux, and freshwater flux – are transferred from ECHAM5 to NEMO; while six variables – including sea surface temperature, fraction, temperature, albedo of sea ice, and surface ocean current speed – are transferred from NEMO to ECHAME5 (Fig. 1). The vector of wind stress is passed to NEMO for two grids – the u grid and v grid of the ORCA2 mesh. Bilinear interpolation is used to re-grid variables exchanged between NEMO and ECHAM5 grids during the transfer process in OASIS3. The vector variables are rotated during transfer because the grid directions are different in NEMO and ECHAM5. The time step of both the atmospheric and oceanic model is 2400 seconds. The coupling frequency is once every 4 hours (six time steps).

The OASIS coupler on the whole needs little modification if we purely follow the COSMOS data exchange procedure; however, in order to precisely control the coupler model to stop and restart at any step by the unit of 'step', not the unit of 'seconds', some patches needed to be added.

**2.4 Support libraries**

The COSMOS coding system placed some common subroutines into the support library. The internal support library directories can be found in the location 'src/lib'. The first group of libraries are used for OASIS interpolation (include anaisg, anaism, fscint, and scrip) and OASIS communications (including clim, mpp_io, and psmile). The second group is for supporting ECHAM5 (including blas, lapack, and support), and the third group is for NEMO, which is named IOIPSL. Moreover, some external libraries are needed, such as NETCDF and MPI.

The IOIPSL library was ported from outside of COSMOS. Different versions of NEMO are compatible with different versions of the IOIPSL library; for example, NEMO2.3 requires IOIPSL-v2. As mentioned, the sources of IOIPSL are located in the 'lib/ioipsl' directory, and one should first edit the file defprec.f90 to define the precision levels of the output data. The current options that can be applied are '–i4 –r8', which means output the 32-bit integer and double the precision of the float-point data. The compiled binary file is named libioipsl.a, which should be placed in the object directory for linking. In addition, some of the modules ('.mod' files) of IOIPSL need to be copied to the directory nemo/include for NEMO compilation.

The external library NETCDF is commonly used for many climatic models. Here, we only need version 3.6.0 or later. It can be compiled by the gcc or icc compiler in X86_64 format modules. In order to support 2GB and larger file input/output, one must enable the option to support 64-bit offset data when compiling NETCDF.

ICM requires the MPI communication library, which supports the MPMD program execute model, and the MPICH2 or Intel MPI are both suitable. Because the MPI library always connects with high-speed interconnection hardware, different manufacturers thus provide different preferred MPI libraries. In our opinion, at the model development stage, the Intel MPI is better for ICM. When the model is used for long-term simulation or operational use, one needs to select the manufacturer's preferred MPI library at that time.

**2.5 Compiler and build tools**

The COSMOS model system was originally built in the NEC-SX6 vector computer system. However, at our institute and the National Supercomputer Center in Tianjin, we use a Linux cluster as the developmental and experimental platform. This PC-cluster runs on an X86_64 CPU processor provided by Intel (or AMD, compatible with INTEL), and thus the Intel FORTRAN compiler is preferred.

The steps to compile ICM are separately performed by several scripts. Before compiling the component model, users should make sure that the support libraries work well. The script in util/COMP_libs.ksh will generate all the internal libraries and put the objective in the x86_64/build library.

The sequence for compiling the component models is unimportant, and the model compile scripts are named COMP_ECHAM_ICM.ksh, COMP_NEMO_ICM.ksh, and COMP_OASIS.ksh. One can compile ECHAM, NEMO and OASIS individually, and the executable files are locate in the x86_64/bin directory. After successfully compiling all of the three components, ICM can be run as a coupled model, the procedure for which is detailed in section 5.

## 3. Field exchange and data interpolation

### 3.1 ECHAM5 field organization

Most of the coupling codes are in program mo_couple.f90. Some of the data arrays in ECHAM5 are the same as those that handle MPI-OM data, while others are new arrays that handle data from NEMO.

We list the six variables and their corresponding data arrays inside ECHAM in Table 2. The codes that receive fields are placed in the subroutine couple_get_o2a at the beginning of stepon integration loops. The subroutine couple_get_o2a calls prism_get in the prism library to actually receive data and then calls atm_put_o2a to move data to model arrays which distribute to different processors. At the end of the stepon loops, the subroutine couple_put_a2o is called to send 17 variables to OASIS. The definitions of these variables can be found in Table 3 and the variable descriptions are in Table 6.

**Table 2. ECHAM5-received fields via OASIS from NEMO.**

| ECHAM Recv. No. | OASIS No. | OASIS→ECHAM symbol | Field description | Array name in ECHAM |
|---|---|---|---|---|
| 1 | 1 | SISUTESU | Sea surface temperature | tsw |
| 2 | 2 | SIICECOV | Sea ice area fraction | seaice |
| 3 | 3 | SIICETEM | Surface temperature over sea ice | tsi |
| 4 | 4 | SIICEALB | Albedo over sea ice | alsoi |
| 5 | 5 | SIUVEOCE | U-velocity longitude direction | ocu |
| 6 | 6 | SIVVEOCE | V-velocity latitude direction | ocv |

Table 3. ECHAM5-sent fields via OASIS to NEMO.

| ECHAM Send. No. | OASIS No. | ECHAM→OASIS symbol | Array name in ECHAM |
|---|---|---|---|
| 1 | 7 | SOZOTAUX | awust |
| 2 | 8 | SOMETAUU | awvst |
| 3 | 9 | SOZOTICX | aiust |
| 4 | 10 | SOMETICU | aivst |
| 5 | 11 | SOZOTAUV | awust |
| 6 | 12 | SOMETAUY | awvst |
| 7 | 13 | SOZOTICV | aiust |
| 8 | 14 | SOMETICY | aivst |
| 9 | 15 | SOPEFWAT | awfre |
| 10 | 16 | SOPEFICE | aifre |
| 11 | 17 | SOTOSPSU | atsno |
| 12 | 18 | SOICEVAP | aieva |
| 13 | 19 | SOSWFLDO | awsol |
| 14 | 20 | SONSFLDO | awhea |
| 15 | 21 | SOSHFLIC | aiswf |
| 16 | 22 | SONSFLIC | ailwf |
| 17 | 23 | SODFLXDT | adqdt |

## 3.2 NEMO field organization

When using OASIS as a coupler, the code to handle field exchange is in the program flx_oasis_ice.h90. The subroutine 'cpl_prism_send' and 'cpl_prism_recv' organize the six variables for sending and the 17 variables for receiving, respectively. A flow chart demonstrating the field exchange for NEMO is presented in Fig. 3.

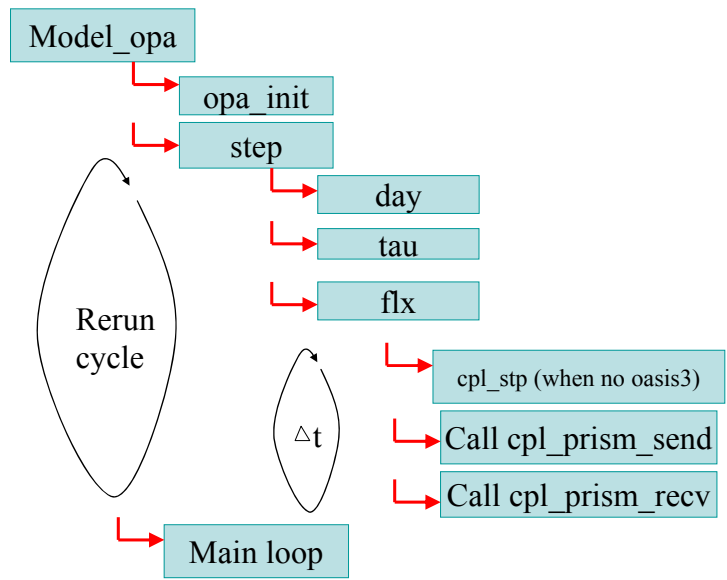

Figure 3.　Flow chart of NEMO.

Table 4 lists the six variables and their corresponding data arrays inside NEMO, and the 17 variables received from OASIS are listed in Table 5. The descriptions of these 23 variables can be found in Table 6.

**Table 4. Fields that NEMO sends via OASIS to ECHAM5.**

| NEMO Send. No. | OASIS No. | NEMO→OASIS symbol | Field description | Array name in NEMO |
|---|---|---|---|---|
| 1 | 1 | SOSSTSST | Sea surface temperature | tn(:,:,1)+rt0 |
| 2 | 2 | SOICECOV | Sea ice area fraction | 1.0-frld |
| 3 | 3 | SOICETEM | Surface temperature over sea ice | tn_ice |
| 4 | 4 | SOICEALB | Albedo over sea ice | alb_ice |
| 5 | 5 | SOUVEOCE | U-velocity longitude direction | un_send |
| 6 | 6 | SOVVEOCE | V-velocity latitude direction | vn_send |

**Table 5. Fields that NEMO receives via OASIS from ECHAM5.**

| NEMO Send. No. | OASIS No. | OASIS→NEMO symbol | NEMO Send. No. | OASIS No. | OASIS→NEMO symbol |
|---|---|---|---|---|---|
| 1 | 7 | COZOTAUX | 10 | 16 | COPEFICE |
| 2 | 8 | COMETAUU | 11 | 17 | COTOSOPR |
| 3 | 9 | COZOTICX | 12 | 18 | COICEVAP |
| 4 | 10 | COMETICU | 13 | 19 | COSWFOCE |
| 5 | 11 | COZOTAUV | 14 | 20 | CONSFOCE |
| 6 | 12 | COMETAUY | 15 | 21 | COSHFICE |
| 7 | 13 | COZOTICV | 16 | 22 | CONSFICE |
| 8 | 14 | COMETICY | 17 | 23 | CODFLXDT |
| 9 | 15 | COPEFWAT | | | |

**3.3 Interpolation method and weighting data**

As indicated, 23 variables are exchanged between ECHAM5 and NEMO, but each of the variable exchanges is not direct because they use different coordinate systems.

As shown in Table 6, the data transferred from NEMO to ECHAM (No. 1–6) moves from an irregular grid to a T31 Gaussian grid via data interpolation. The operation method in OASIS includes MOZAIC (meaning mask) and REVERSE, which changes the data array sequence in the meridional direction.

Table 6. OASIS's operation of fields (E and N denote ECHAM and NEMO, respectively).

| OASIS No. | Data direction | Field description | Operation |
|---|---|---|---|
| 1 | N→E | Sea surface temperature | MOZAIC REVERSE |
| 2 | " | Sea ice area fraction | " |
| 3 | " | Surface temperature over sea ice | " |
| 4 | " | Albedo over sea ice | " |
| 5 | " | U-velocity longitude direction | " |
| 6 | " | V-velocity latitude direction | " |
| 7 | E→N | Zonal wind on water, ugrid | INVERT MASK EXTRAP MOZAIC |
| 8 | " | Meridional wind on water, ugrid | " |
| 9 | " | Zonal wind on ice, ugrid | " |
| 10 | " | Meridional wind on ice, ugrid | " |
| 11 | " | Zonal wind on water, vgrid | " |
| 12 | " | Meridional wind on water, vgrid | " |
| 13 | " | Zonal wind on ice, vgrid | " |
| 14 | " | Meridional wind on ice, vgrid | " |
| 15 | " | P - E on water | " |
| 16 | " | P - E on ice | " |
| 17 | " | Total snow (eau + glace) | " |
| 18 | " | Evaporation on ice | " |
| 19 | " | Solar flow on water | " |
| 20 | " | Non-solar flow on water (LW+sens+lat) | " |
| 21 | " | Solar flow on ice | " |
| 22 | " | Non-solar flow on ice (LW+sens+lat) | " |
| 23 | " | Non-solar heat flow derivative | " |

The data transferred from ECHAM to NEMO (No. 7–23) moves from a Gaussian grid to an irregular grid via interpolation. Because the atmospheric grids might correspond to land when mapped to the ocean grid, the mask operation is need. More details can be found in the OASIS manual [3].

**4. Bugs fixed in the component models and coupled model**

Tuning model performance is the most time consuming work after porting model components to ICM and coupling them. Some bugs can develop when using ICM to perform climate simulations, and in this section we describe these bugs and the solutions to the most major problems.

**4.1 The I8 date problem**

The date in the model is generally recorded by the second, and this date counter system works well for weather predictions. However, the climatic model using this time counter system might experience some problems.

The I8 integer data in Fortran refers to eight hex bytes of data, for which the maximal value is $2^{32}$ if it is unsigned, and half that (i.e., $2^{31}$) if it is signed. OASIS reads a variable $RUNTIME from the Namelist file, and this value is a signed integer to express the number of seconds for which the coupler has run. We know that 1 day has 86400 seconds, and $2^{31}$ seconds thus approximates to 68 years. Therefore, the time counter in OASIS will overflow if the simulation is longer than 68 years in a single run. This limitation means the ICM cannot be run for 100 years without modification.

To deal with this problem, there are two possible solutions. First, one can change the time counter units in OASIS from seconds to steps, which is the situation in ECHAM and NEMO. Since one step is about 2400 seconds, the step counter system can allow a simulation up to 163,000 years long. However, such a lengthy single run carries some disadvantages. For example, a 500-model-year simulation would need 5 to 10 days to complete, and during this times the machine may encounter faults, meaning experiments will stop midway without outputting any restart information. The experiments would need to be rerun from the beginning, and this would waste a great deal of CPU time if the simulation crashed near the end of a 500-model-year run.

Therefore, we prefer the second solution to enable simulations of longer than 68 model years. This involves separating the 500 model years in to dozens of 10-model-year runs, with each run satisfying the I8 time limit (i.e., runs of fewer than 68 model years). Because we can make the ICM restart the run output with the same binary data as a single run, the two approaches to solving the I8 date problem are equivalent.

**4.2 The NEMO perfect restart**

Long-term ICM simulations are generally best separated into a series of short-term runs, e.g., 10 years for each experiment. However, when we compared the 11–20 year output data with the 11–20 year data from another 20-year experiment, we found they were different. In order to find a solution to this problem, we ran the component model in the uncoupled mode. The results indicated that ECHAM5 outputs the same result when uncoupled, and thus the problem is routed in the NEMO model or in the coupling process.

We set up and ran NEMO in the uncoupled mode, and we noticed that NEMO indeed cannot output the same data for a restarted 11–20 year run and the corresponding data from a 20-year experiment. Following this, we carefully checked the code, and found some bugs. First, the compile option '-Dkey_vectopt_loop' is a preferred option in NEC vector computers such as NEC-SX6, which is commonly used in Japan and Europe. However, this key will cause data array overflow when coded to compile in a Linux system.

Second, the MPI parallel scheme in NEMO2.3 is not robust, and it does possess some problems when we define '-Dkey_root_exchg' to exchange data by the root process in NEMO. This bug is not triggered when the model runs within 1 CPU. However, when multiple CPUs are used and the program calls the cpl_prism_recv or cpl_prism_send subroutines, a global data array will lose some data if the i-direction CPU number is larger than 1. We applied a patch to fix this bug to allow NEMO2.3 to perform two direction (i and j) in parallel at the same time. Another small bug is that tradmp.F90 stores a data array, but indexes it from No. 2. Thus, the first data in the array are random, and this will cause the output data to produce different results. We set the first data in the array to zero, and then the results were identical.

The ice model is integrated in OPA, but the restart file for the ice model can only be written at a step that is exactly divisible by five. Therefore, if we need to stop the ICM model for a whole day (which corresponds to multiples of 36 steps), it will not output the ice model data correctly. This limitation was avoided after we modified the ice model code.

Ultimately, we repaired all of the above-mentioned bugs, and a test run indicated that the patched version of NEMO2.3 can output consistent binary results for both a restarted run and a direct run in the uncoupled mode.

**4.3 The timing sequence for component model communication**

The two component models in ICM were both restarting successfully when the model was run in the uncoupled mode; however, they were still not producing consistent binary output data under the coupled run. We analyzed the problem and found that the timing sequences of communication were important in the restarted run.

The timing sequences in ECHAM5 can be obtained by enabling the debug message when compiling with the option '-D__synout', and the debug information is written into the file 'atmout', which is located in the experiments directory.

The steps by which ECHAM exports data to NEMO are that the stepon calls couple_put_a2o at the steps that the residue is 5 when isteps divide by 6. Every communication step outputs 17 groups of data. The variable istep denotes the step number, and when istep%6=0,1,2,3,4, the couple_put_a2o returns directly and does nothing. The data are sent to the ocean model when istep%6=5. For example, we set the ICM restart at 72 steps (2 model days), and istep=71 was the last time to call couple_put_a2o in this run. The data were received by OASIS, and the OASIS time counter system found these to be the last data sent to NEMO, but they should not have been sent out and written to the OASIS restart file (flxatmos).

The steps by which ECHAM imports data from NEMO is handled by couple_get_o2a, is activated at istep%6=0, and each time six groups of data are received. The first step is when the restart run begins (istep=0), and at this moment the data are received from the restart file via OASIS. The second data receive date is when istep=6, which is the time the data are actually received from the ocean model. When we select 72 as the end step, the last step that ECHAM receives data from OASIS is istep=66 because istep=71 is when the rerun files are written. The run continues to restart from istep=72, but the data in the first step are from the file sstocean via OASIS.

The binary file flxatmos contains variables COZOTAUX etc., which contains a total of 17 variables and the data size is 627096 (96×48×17×8) bytes. Meanwhile, sstocean contains a total of six variables and the data size is 1301808 (182×149×6×8) bytes. At the beginning of the restart run, ECHAM5 first reads the rerun file (called iorestart in control.f90) and then receives data from OASIS. Note that the rerun generating subroutine is in stepon.f90 (called savehis).

The timing sequences in NEMO can be obtained by enabling the debug message by editing the namelist file and setting 'ln_ctl=.true.', and the debug information is written in the files named ocean.output_0001. NEMO exports the date to ECHAM5 by calling cpl_prism_send at the step satisfying date%6=5 (where the variable 'date' in NEMO means 'step'), and each time six groups of data are sent. The output frequencies are the same as in ECHAM5, and the steps (date%6=0,1,2,3,4) in the cpl_prism_send subroutine do nothing. In the 2-days restart run example, the step (date=71) represents the last export of data; and after the data have arrived in OASIS, the coupler does not send out any further data and writes them to the restart file flxatmos.

NEMO imports data from ECHAM5 by calling cpl_prism_recv at the step satisfying date%6=0. In the 2-day restart run example, at step date=66 the data are received from OASIS for the last time, and at step date=72 the rerun file for NEMO is generated. The continued run of NEMO restarts from nit000=73, but the data in the first step are from flxatmos via OASIS.

A notable difference to ECHAM is that the ICE model is integrated in NEMO, and the internal coupling between ICE and OPA is controlled by a parameter nfice in the namelist file. We set nfice=6 in the current version, which means that every six ocean steps the ICE model is called once.

In order to debug ICM, we sometimes needed to look at the OASIS information, and this information was outputted to the file cplout when the 'nlogprt' parameter was larger than two in the file namcouple.

OASIS handles different files by file number; for example, sstocean and flxatmos correspond to No. 21 and No. 27, respectively. The variable file number mapping array, nluinp, begins from 21, and thus the first variable corresponds to the No. 21 file (corresponding to sstocean). The first six file numbers in nluinp are 21 and the 7th to the 23rd variables map to the file number 27 (corresponding to flxatmos).

The first cycle when OASIS starts up is the step control variable kiter=0, and this time the coupler reads the restart data from files. In later cycles, data are received from ECHAM and NEMO individually by calling CLIM_import. After the interpolation has been performed, the data are sent out by calling CLIM_export.

We carefully tuned the timing sequence for the three relative model components and ultimately were able to make ICM perform a perfect restart in the coupled mode.

**5. The scripts system**

ICM was designed to be a hybrid model for research and operational application, and the ICM directory structure contains development directories such as 'src' and 'lib', and an operational directory called CMSR_ICM. In this section we describe how to set up and run an experiment in ICM. A flow-chart summary for running a simulation case is presented in Table 7.

**Table 7. Flow chart for running a simulation case.**

| No. | Step description |
|---|---|
| 1 | Make a directory 'case_dir' |
| 2 | Copy initial data files |
| 3 | Copy model executable files |
| 4 | Copy namelist files |
| 5 | Copy job_file |
| 6 | Submit job |
| 7 | If finished successfully, save restart |

|                     | file                |
| ------------------- | ------------------- |

## 5.1 Data preparation

The first task that a job script does is to prepare the working directory and copy the initial or restart data for simulation.

Table 8 shows some of the major settings in the scripts. For example, the case name here is 'ctl-1500', and the scripts will try to read a file located in CMSR_ICM/ctl-1500/ctl-1500.trace to find out if the case should be an initial restart run. If the file does not exist, the scripts start the new case procedure and copy the initial file to that directory (Tables 9–11). Otherwise, the script will copy the saved restart file by analyzing the date in ctl-1500.trace to the working directory.

**Table 8. Predefined settings in run-th.sh.**

| No. | Parameters and settings | Description |
| --- | --- | --- |
| 1 | CEXPER=ctl-1500 | The case name 'ctl-1500', in the top of CMSR_ICM |
| 2 | NEMO=/disk1/user1/gcm/ICM/data/ini/NEMO DATAB=/disk1/user1/gcm/ICM/scripts SCRATCHB=/disk1/user1/gcm/ICM/scripts HOMEB=/disk1/user1/gcm/ICM/scripts | Data path |
| 3 | NPROCA=8 NPROCB=4 | The CPUs in ECHAM5, totally NPROCA× NPROCB |
| 4 | NCPUS_NEMO=32 | The CPUs in NEMO, precompiled binary oceanx-32cpu should exist |
| 5 | YEAR_BEGIN_EXP=1 MONTH_BEGIN_EXP=1 DAY_BEGIN_EXP=1 | The date for initial run |
| 6 | CPEXP_DUR=$(( 18000 )) | Total months for the case (1500 yr=18000 mon) |
| 7 | JOB_DUR=120 | Months for each restart run (10 yr) |
| 8 | MAXSUBJOB_DUR=120 | Months for each restart run (10 yr) |

**Table 9. Initial data for ECHAM5 (T31).**

| No. | Variable description | File name |
| --- | --- | --- |
| 1 | Hydrological parameters | hdpara.nc |
| 2 | Hydrological initial data | hdstart.nc |
| 3 | Climatic SST (unused when coupled run) | sst1977 – sst2000 |
| 4 | Climatic SIC (unused when coupled run) | ice1977 – ice2000 |
| 5 | Surface temperature (AMIP2, monthly data) | unit.20→T31/T31_amip2sst_clim.nc |

| 6 | Zonal mean ozone climatology (monthly data) | unit.21→T31/T31_O3clim2.nc |
| 7 | Spectral 3-dim initial file of the atmosphere | unit.23→T31/T31L19_jan_spec.nc |
| 8 | Surface boundary conditions | unit.24→T31/T31_jan_surf.nc |
| 9 | Annual cycle of the leaf area index | unit.90→T31/T31_VLTCLIM.nc |
| 10 | Annual cycle of vegetation ratio | unit.91→T31/T31_VGRATCLIM.nc |
| 11 | Annual cycle of Land surface temperature | unit.92→T31/T31_TSLCLIM2.nc |
| 12 | Surface temperature over ICE | unit.96→T31/T31_amip2sic_clim.nc |
| 13 | Input for radiation scheme | rrtadata→ T31/surrta_data |

**Table 10. Initial data for NEMO.**

| No. | Variable description | File name |
| --- | --- | --- |
| 1 | Net freshwater budget | EMPave_old.dat, STRAIT.dat |
| 2 | Zonal eddy coefficients west of equator | ahmcoef |
| 3 | Terrain data at bottom of sea | bathy_meter.nc, bathy_level.nc |
| 4 | Coordinates at grid | coordinates.nc |
| 5 | Monthly climatic temperature | data_1m_potential_temperature_nomask.nc, |
| 6 | Monthly climatic salinity | data_1m_salinity_nomask. nc |
| 7 | Heat flux | geothermal_heating.nc |
| 8 | Monthly climatic runoff | runoff_1m_nomask.nc |
| 9 | Monthly climatic wind stress | taux_1m.nc, tauy_1m.nc |

**Table 11. Initial data and interpolation data for OASIS.**

| No. | File description | File name |
| --- | --- | --- |
| 1 | Description of the exchange variable names | cf_name_table.txt |
| 2 | Atmospheric fluxes | flxatmos |
| 3 | Ocean surface conditions | sstocean |
| 4 | Longitudes and latitudes of all grids in the coupling | grids.nc |
| 5 | See land masks of all grids involved in the coupling | masks.nc |
| 6 | Grid cell surfaces for all grids involved in the coupling | areas.nc |
| 7 | Interpolate ocean grid to T31 grid | orcTt31, orcUt31, orcVt31 |
| 8 | Interpolate T31 grid to ocean grid | t31orcT, t31orcU, t31orcV |
| 9 | Weights and addresses used for extrapolation | gweights, mweights, nweights |

Figure 4 shows the structure of the job working directory, where the initial data are copied from the data directory to the ctl-1500/run directory.

```
CMSR_ICM
|-- Namelists
|-- bin
|-- scripts
`-- ctl-1500
   |-- bin
   |-- input -> ../../data/input
   |-- log
   |-- out
   |   |-- echam5
   |   `-- nemo
   |-- restart
   |-- run
   |   `-- out
   `-- scripts
```

Figure 4.     Experiment tree of ICM.

## 5.2 Namelist file generation

In order to provide flexible control of the simulation case restart, the namelist file for the three models are auto-generated by scripts. The basic namelist files are stored in the CMSR_ICM/Namelists directory including amatmos.base, namcouple.base, namelist.base, and namelist_ice. Inside the script run-th.sh, ksh is used as an interpreter to perform some arithmetic operations, and thus the software /bin/ksh or /usr/bin/ksh must be installed beforehand. The dates that the model output files apply are based on the Gregorian calendar system, and the computation needs two external tools, caldat and julday, to work.

We demonstrate here the namelists generation for an initial ICM run. The namelist for NEMO is run/namelist; we set 'ln_rstart = .false.', 'nrstdt= 0', 'nit000= 1', and 'nitend=13140' for a one-model-year simulation. The namelist file for ECHAM5 is run/namelist.echam, and 'LRESUME=.false.' indicates an initial run. The options DT_START, DT_STOP and PUTRERUN control the model simulation time. When we need to restart by one year the 'PUTRERUN= 1,'years','first',0' is set. OASIS opens run/namcouple as namelist, and $RUNTIME and $INIDATE are needed to be set the correct values. For example '$RUNTIME =31536000' when

starting from '$INIDATE =00010101' and stopping at the end of 1 model year.

**5.3 CPU resources**

The CPU number used by ECHAM5 and NEMO are indicated by No. 3–4 in Table 8. ECHAM5 is compiled as a single executable file that can perform different CPUs in parallel, and the total computation process is determined by NPROCA× NPROCB. Changing the CPU number need not rebuild the binary file.

However, the parallelization scheme in NEMO defines the CPU allocation in compilation time, not execution time, and thus changing the CPU numbers for NEMO would mean rebuilding it every time. To overcome this shortcoming, we pre-compiled some commonly used executable binary files and put them in the 'bin' directory for later use. For example, the 32-CPU NEMO model is stored as ocean-32cpu, and so on.

Currently, one CPU is used in OASIS. The script reads three CPU numbers for the three models and then calculates and generates the CPU numbers in the namcouple file. Meanwhile, the MPI executes a command line corresponding to these three CPU settings that are generated.

**5.4 Job restart**

For restarting a job, a script will generate the restart option in the namelist files. The changed items in the NEMO namelist file are 'ln_rstart = .true.' and 'nrstdt=2'. The steps that NEMO follows to restart from and end with are calculated by an external program, julday, and the item nit000 is the nitend value of the last run and plus 1. The item nitend is set to the end days of this run and multiplied by 36 steps.

The restart data files for NEMO are in two series: one is for OPA and the other series is for the ICE model. The filenames for the OPA restart data are in the form ORCA2_00011231_restart_0000.nc, which contains the stop date and the parallel process ID within. The ICE model restart data are in the form ORCA2_00011231_restart_ice_0000.nc, are placed in the ctl-1500/restart directory, and renamed to restart_0000.nc and restart_ice_in_0000.nc, respectively. The script automatically copies them back to the working directory for the continued run.

For restarting the ECHAM5 model, the item 'LRESUME= .true.' is set in run/namelist.echam, and the other items in that file need not to be modified. ECHAM5 reads spectral data from the rerun_CTL_echam file, which is stored in the ctl-1500/restart directory. Since at the end of the ECHAM5 simulation the old rerun_CTL_echam file will be overwritten, we must thus copy them to different directories for later use. If the hydrology model is activated, hdrestart.nc should also be kept.

The restart files for OASIS are flxatmos and sstocean, and they are modified at the end of the simulation and stored in ctl-1500/restart. We need to copy them back to the working directory. The step control in OASIS is different to ECHAM and NEMO because the date size is not cumulative.

After the three namelist files are generated, the run scripts call the MPI start program and executes it, which is the same as the command that the initial run has performed.

**5.5 Post data processing and re-grid tools**

The output data for ICM include atmospheric and oceanic data, and generally a long-term climatic simulation will generate thousands of files when monthly data are saved. In addition, ECHAM5 outputs the spectral coefficient in the data for improving computing performance, and NEMO outputs local data in each process. Thus, all of these data need to be post processed for later analysis.

Owing to ECHAM5 providing a data preparation tool named 'after', by applying this tool the user can select the variables as well as the level and generate Guassian grid data from spectral coefficient outputs.

NEMO data will be in the form of something like ORCA2_1m_SSSS_EEEE_grid_T_CCCC.nc, where 'SSSS' denotes the start time, 'EEEE' the end time, and 'CCCC' the process ID. When executing NEMO in parallel, for example using 32 CPUs, the 32 separate files should be merged first by the tool 'flio_rbld', which is ported from the package modipsl.

It should be noted that the irregular coordinate system that NEMO applies cannot be handled well by some graphics software, such as GrADS and NCL [19]. However, Ferrete can plot them correctly. To apply NCL to analyze the data, we provide re-gridding tools to interpolate the data to a general regular grid.

In addition, we also provide a basic diagnostic plots package to evaluate the model simulation result. This can draw figures, as shown in Huang et al. [5], as well as some other diagnostic plots.

**6. Computational performance**

High-performance computers or supercomputers with tens of thousands of cores present the opportunity to develop high-performance models [20-22]. The latest supercomputers have a large amount of cores installed, enormous storage, and rapid point-to-point communication. At the time of writing, the most powerful supercomputer in the world is China's Tianhe-2 computer[1], achieving an impressive 33.86 P-flops per seconds using 32000 Intel Xeon E5-2692 (12-core 2.200GHz) and 48000 Xeon Phi 31S1P processors. In China, the second best supercomputer is TH-1A with a 2.57 P-flops performance level, over 186,368 processing cores, 229.4 TB of memory, and a 1 PB storage system. Since the Tianhe-2 computer will not be open to outside users until late 2014, we perform our experiments in the TH-1A machine, but the two systems are very similar, meaning the model can be easily ported to Tianhe-2 once it goes into service.

**6.1 IO and debug message**

The IO speed and bandwidth are very important for large-scale and high-speed climate models in a cluster-based supercomputer[23, 24]. A benchmark study was carried out by LASG and Tinghua University, which indicated that the IO may cost about 90% in all simulation times in the TH-1A system for very high resolution climate models[2] when the CPU cores are larger than 5000.

---

[1] http://top500.org/lists/2013/11

There is a lot of debug information as a result of tuning the model performance, as indicated in section 4.3. Much of this information is written to files in disks, and thus when performing long-term simulations the files will waste a large amount of disk space and, moreover, the extra IO will decrease the model speed. For this reason, there is a need to disable the unnecessary debug information IO at the compilation and execution stages.

The debugger file for OASIS is stored in cplout, and it is controlled by NLOGRT in the namcouple file. Setting it to a negative value will greatly reduce the size of cplout. There are two files in NEMO named ice.evilu and ocean.output that diagnose the ICE and OPA status every few steps. These can be disabled by deleting the write code in limdia.F90 and setting a parameter 'lwp=0' in the namelise file.

Debug messages for ECHAM5 are added when compiling with the option '-D__synout', so deleting this option and rebuilding the executable binary files will reduce the IO frequency in ECHAM5.

**6.2 CPU number matching**

As indicated, the ICM runs in MPMD mode, which means three different sets of executable files run at the same time. The communication timing sequence between ECHAM and NEMO was analyzed in section 4.3. The models export data, directly return to the main loop, and then in the next step import data from OASIS. The import procedure is not directly returned and waits until the data has arrived. Therefore, sometimes the ECHAM is waiting for NEMO and sometimes NEMO is waiting for ECHAM. To improve the speed of simulation, we need to evaluate the time cost between one import calling another and try to reduce the waiting times.

The time cost in a single importing cycle is dependent on the computer speed, i.e.,

---

[2] Personal communication with Dr. Liu Hailong and Xue Wei, 2013.

the power of the CPU. In Tables 12–13 we list some effective CPU allocation schemes for running ICM in the TH-1A computer.

**6.3 High-speed interconnection effect**

The faster the component models, the faster the speed that can be gained for ICM. ECHAM and NEMO apply MPI parallelization internally, and the communication speed between cluster nodes is very important. We compared a Dawing cluster with a 10GB IB card and the TH-1A computer, which uses 80GB internal connections, and found that when the computation scale is larger than 32 CPUs, the TH-1A can provide more efficient (about five times faster) simulation, mainly due to the higher interconnection speed.

A notable influence on the simulation speed that is sometimes overlooked is that of the cache and memory. Because climatic models – especially high-resolution models – cost a lot of memory, the cache rate decreases if all the cores are participating in the computation. For example, in TH-1A we use 10 cores in each node (total of 12 cores) and gain 50% more speed than when all 12 cores are used. The two free CPUs in the nodes are not involved in the computation but can still be used in MPI communication and perform IO jobs for the system. Therefore, preserving 1–2 CPUs in cluster nodes seems beneficial in climate simulations.

**6.4 Speed benchmark of ICM**

Having dealt with many issues to improve the ICM's simulation speed, we now demonstrate the performance it achieves in TH-1A. From the information in Table 12, we can conclude that ICM can obtain 115 model years per day by applying 16 ECHAM processes and 32 NEMO processes. In addition, when the processes in NEMO are kept at 32 CPUs, the time increase from 16 to 32 ECHAM processes is small. This means that the 32 processes of ECHAM are indeed waiting for NEMO to finish its computation most of time, and so increasing the speed of NEMO is necessary. Finally, when the CPUs for NEMO are increased from 32 to 64, the new speed synchronization established and the global speed increases from 115 yrs/d to 175 yrs/d.

Table 13 lists the simulation speeds for a different resolution of ECHAM (T63) than that represented by Table 12. Since the resolution increase and the step size for T63 is half that used by T31, the speed of ICM is not as fast as the low-resolution version. However, we can still perform 60–70 yrs/d simulations in TH-1A.

**Table 12. ICM speed in the TH-1A supercomputer with T31 ECHAM5 resolution.**

| No. | seconds/model-day | model-years/day | ECHAM CPUs | NEMO CPUs |
|---|---|---|---|---|
| 1 | 3.75 | 63 | 8 | 16 |
| 2 | 2.05 | 115 | 16 | 32 |
| 3 | 2.00 | 115 | 32 | 32 |
| 4 | 1.35 | 175 | 32 | 64 |

**Table 13. ICM speed in the TH-1A supercomputer with T63 ECHAM5 resolution.**

| No. | seconds/model-day | model-years/day | ECHAM CPUs | NEMO CPUs |
|---|---|---|---|---|
| 1 | 3.20 | 71 | 64 | 16 |
| 2 | 3.90 | 58 | 64 | 32 |
| 3 | 3.11 | 71 | 128 | 32 |
| 4 | 4.10 | 56 | 64 | 64 |
| 5 | 3.00 | 75 | 128 | 64 |

**6.5 Long-term control simulation and ENSO ensemble hindcast**

In order to evaluate ICM's performance, a 1500-year control run has been performed. ECHAM has outputted 18,000 files totaling 70GB, while NEMO outputted 24,000 files with a total size of approximately 1TB. This control run was restarted every 10 model years, and all of the restart files were also stored for future use. With the help of TH-1A, the job took about half a month to complete.

We also applied ICM to generate initial perturbation data for ENSO simulation, including hindcast and forecasting. The 6–10-group ensemble simulations indicated the model can simulate ENSO well. For more details, readers are referred to Wang et al. [6]. Moreover, the daily output data of ECHAM5 model for some short term simulation (for example 1 model-year) can be used to analysis the coupled model's uncertainty due to computation conditions[25].

## 7. Summary and perspective

This manuscript describes the software engineering and computational performance of Integrated Climate Model (ICM-1.0) atmosphere–ocean general coupled model applied by CMSR/IAP. ICM uses ECHAM5 and NEMO 2.3 as its atmospheric and oceanic components, respectively, coupled by OASIS3. The first version of ICM is a low-resolution version, with a horizontal resolution of T31 or T63 for the atmospheric component and around 2° for the oceanic component. The model has steadily integrated up to 1500 years without climate drift. Evaluations [5] have shown ICM successfully simulates the main characteristics of interannual variation of tropical Pacific SST and the EA–WNP summer monsoon, as well as reproduces their connection mechanism.

The organization of ICM model's code and the definition of coupling fields are described. Model components porting and bug fix also demonstrated. The improvement of ICM's coding is that we enabled it to perform perfect restart runs, which is an important feature that was not implemented in the original version of NEMO2.3. We apply scripts to manage the task job and allocate CPU to do effective simulations in Dawning and TH-1A system.

The ICM model has high computational performance due to the highly efficient parallel scheme applied by the component models. The fully coupled simulation speed in TH-1A can reach 100 model years per day (maximum of 175 yrs/d). This advanced and compliant COSMOS model system enable users to DIY your own new climate model or change component models based on ICM.

**Acknowledgements.** This research was jointly supported by the National Natural Sciences Foundation of China (41375112) and the National Basic Research Program of China (2011CB309704). The author thanks Dr. J.R. Jiang for discussion on model restart coding, and also thanks to B.L. Yan, P. Huang, K.M. Hu, Y. Liu, G. Huang and X.F. Cui for provide model evolution results and plotting tools. Dr. Q. Bao helps to achieve the performance test data in TH-1A supercomputer.